# Ultrafast electron localization and screening in a transition metal dichalcogenide

Z. Schumacher[1], S. A. Sato[2,3], S. Neb[1], A. Niedermayr[1], L. Gallmann[1,*], A. Rubio[3,4], U. Keller[1]

[1]Department of Physics, ETH Zürich, 8093 Zürich, Switzerland

[2]Center for Computational Sciences, University of Tsukuba, Tsukuba, Ibaraki 305-8577, Japan

[3]Max Planck Institute for the Structure and Dynamics of Matter and Center for Free Electron Laser Science, 22761 Hamburg, Germany

[4]Center for Computational Quantum Physics (CCQ), Flatiron Institute, New York 10010, USA

*Corresponding author. gallmann@phys.ethz.ch

**The coupling of light to electrical charge carriers in semiconductors is the foundation of many technological applications. Attosecond transient absorption spectroscopy measures simultaneously how excited electrons and the vacancies they leave behind dynamically react to the applied optical fields. In compound semiconductors, these dynamics can be probed via any of their atomic constituents. Often, the atomic species forming the compound contribute comparably to the relevant electronic properties of the material. One therefore expects to observe similar dynamics, irrespective of the choice of atomic species via which it is probed. Here, we show in the two-dimensional transition metal dichalcogenide semiconductor MoSe$_2$, that through a selenium-based transition we observe charge carriers acting independently from each other, while when probed through molybdenum, the collective, many-body motion of the carriers dominates. Such unexpectedly contrasting behavior can be traced back to a strong localization of electrons**

**around molybdenum atoms following absorption of light, which modifies the local fields acting on the carriers. We show that similar behavior in elemental titanium metal[1] carries over to transition metal-containing compounds and is expected to play an essential role for a wide range of such materials. Knowledge of independent particle and collective response is essential for fully understanding these materials.**

Understanding ultrafast carrier dynamics is important for the engineering of materials towards applications in electronic and optoelectronic devices. Transition metal dichalcogenides (TMDCs) have sparked great interest in fundamental and applied science due to the abundance of observed physical phenomena [2,3]. Their unique electronic and mechanical properties, allowing to go from an indirect bandgap in bulk material to a direct bandgap by exfoliation to a monolayer, have contributed significantly to their popularity. Many effects, such as the strong exciton binding energy, the long exciton lifetime and the spin selective valley dynamics have been studied extensively with optical and electronic probes. Combining different TMDC monolayers into new heterostructures has introduced further effects such as interlayer excitons and charge transfer [4-6] and trions in Moiré patterned potentials [7]. Besides exfoliation techniques, which allow for precise control of the thickness and stacking of heterostructures, progress in chemical vapor deposition (CVD) has enabled the growth of large-area mono and few-layer samples [8]. Some physics aspects, like dark excitonic states [9], exciton generation time [10] and interlayer charge transfer [11,12] are still actively researched due to the difficulty of characterizing such states. To date most studies focus on the valley, spin, or layer dependent properties, and only few studies have focused on the element specific response of TMDCs [13-15].

Attosecond transient absorption spectroscopy (ATAS) allows for element and carrier specific probe of excited states dynamics and has become a powerful tool to investigate carrier dynamics on the few femtosecond scale in semiconductors [16-20] and metals [1,21], and strong field effects in dielectrics [22,23]. In most of these cases, ATAS uses a broadband XUV pulse to probe the transition between core and excited states with attosecond temporal resolution. With the probe being resonant with a characteristic core level, ATAS becomes inherently element specific while its broad bandwidth simultaneously reveals the dynamics of both, pump-excited electrons and holes.

Here, we apply ATAS to a few-layer CVD grown 2H $MoSe_2$ (6Carbon Inc.) with a bandgap of ≈1.55 eV. Surprisingly, we find qualitatively different dynamics in both conduction and valence band for probe transitions originating from either Mo or Se. This is remarkable because the band structure in the vicinity of the bandgap is formed by covalent bonds between the *d*-orbitals from Mo and the *p*-orbitals from Se with the two contributing to the density of states with similar magnitude. Such a qualitative difference in response was previously observed in the similar material $MoTe_2$ but not explained [13]. In a recent publication, contrasting behavior observed for two W core level transitions in $WS_2$ was suggested to originate from different degrees of localization of the initial states [24]. In contrast, by combining our observed transient spectral features with *ab initio* calculations, we are able to attribute the qualitative difference to pump-induced real-space carrier localization into *d*-orbitals of the transition metal and the resulting local screening modification. Interestingly, the Se response remains entirely unaffected by such many-body dynamics. In $MoSe_2$ we find the localization effects to last longer than in the elemental transition metal Ti and suggest that this faster decay is due to the much higher electron-electron scattering rates in the metal compared to a material with a bandgap.

We simultaneously probe the core level transitions from Mo $4p$ states and the spin-orbit split Se $3d$ states to the valence and conduction band (VB/CB) (Fig. 1). For multilayer MoSe$_2$ the maximum of the valence band is located at the Γ-point and has mostly $4d$ and $4p$-orbital contributions from Mo $d_{3z^2-r^2}$ and Se $p_z$ orbitals, respectively. The orbital character of the conduction band minimum originates predominantly from $d_{xy}$ and $d_{x^2-y^2}$ of Mo and $p_x$, $p_y$ orbital contributions of Se. Furthermore, a non-negligible $d_{3z^2-r^2}$ and $p_z$ orbital contribution from Mo and Se atoms, respectively, is present in the conduction band [25-27]. The strength of the $d$ and $p$-orbital contributions of the two respective atoms to the covalent bonds forming the band structure around the bandgap is comparable as shown in our calculated projected density of state (PDOS) (Fig. 1c).

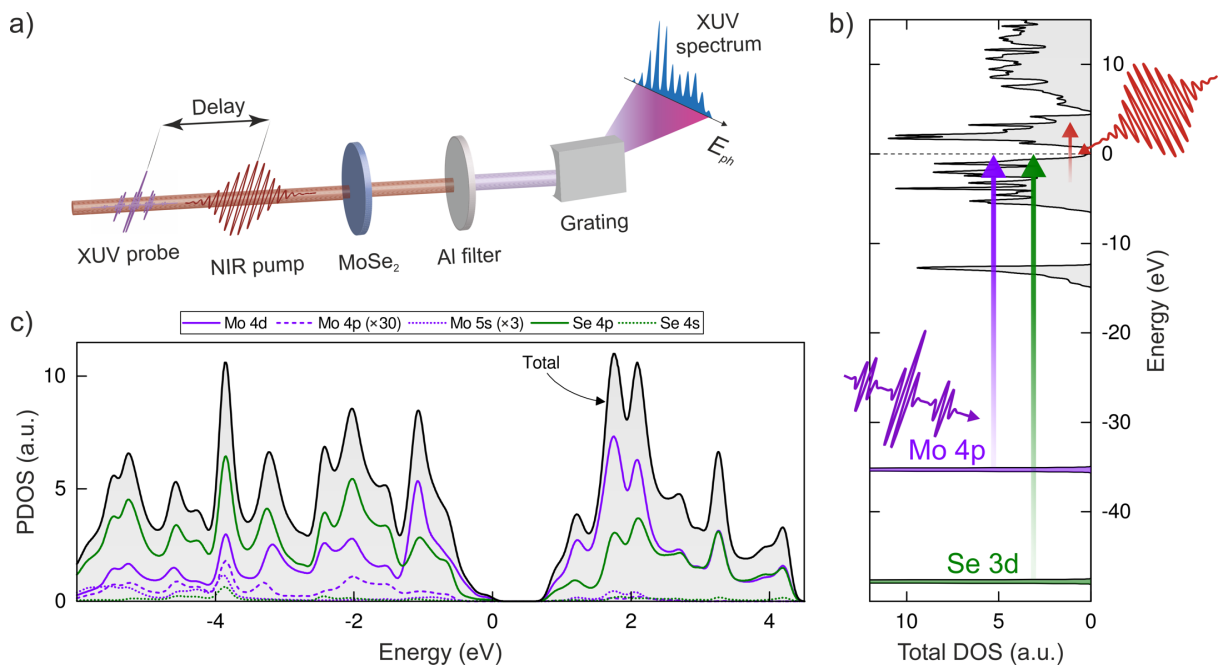

**Fig.1. a)** Experimental setup. A near infrared (NIR) pump and XUV probe pulse are delayed in time. The residual NIR is eliminated by an aluminum filter before the transmitted XUV radiation is measured in a spectrometer. **b)** Total density of states of bulk MoSe$_2$ calculated by density functional theory. The NIR pump excites carriers across the Fermi level, while the XUV probes the transitions from the Mo $4p$ and Se $3d$ core levels to the valence and conduction band (spin-orbit splitting of Se $3d$ not shown). **c)** Projected density of states showing dominant contributions of Mo $4d$ (purple solid line) and Se $4p$ (green solid line) states to the valence and conduction bands. Mo $4p$ and Mo $5s$ have been multiplied by a factor for better visibility.

Our few-layer MoSe$_2$ sample was transferred onto a 30 nm thin silicon nitride membrane as a substrate (Ted Pella) with excellent transparency in the energy window of our experiment. The broadband XUV probe spectrum from an attosecond pulse train spans photon energies from 30 to 70 eV. We use a 10-fs near infrared (NIR) excitation pulse with a center photon energy of 1.54 eV and a peak intensity of $\approx 6 \cdot 10^{11}\,\text{W}/\text{cm}^2$ at 1 kHz pulse repetition rate to promote carriers across the bandgap with a density of $\approx 2 \cdot 10^{21}/\text{cm}^3$. The relatively high pump-induced carrier density can be treated as an electron-hole liquid since it is above the Mott-transition (see SI). The interferometric ATAS setup housed in vacuum is described in more detail elsewhere [28].

**Spectral response**

The pump-induced transient changes in XUV absorption are shown in Figure 2a as a function of time delay between the NIR pump and the broadband XUV probe pulse. Lineouts of that data integrated over a time span of ±20 fs for several time delays are plotted in Figure 2b. The different core level energies allow us to separate the element specific response with the probing photon energy. The Mo response covers the XUV photon energy range between 32 to 50 eV with the strongest signal for the Mo 4p core level to valence and conduction band transition (32 – 43 eV) and some weaker excitations into higher-lying and delocalized states above 47 eV. Of these, the 5s states contribute the most among the bound states (see SI). The Se response for the XUV photon energy range of 54 to 60 eV shows two sharp alternating bands with increase and decrease in absorption that resemble the derivative of an absorption peak in shape. The double appearance of this derivative-shaped feature arises from the spin-orbit split Se $3d_{3/2}$ and $3d_{1/2}$ transitions with an energy difference of ~1 eV (see SI). In comparison the Mo specific signal in the valence and conduction band is more broadband and positive. As is shown below, this drastically contrasting behavior is not a probe-induced effect. Due to the mixed Mo

*d*- and Se *p*-orbital nature of the valence and conduction bands accessed by the XUV probe pulse such drastic difference in response is unexpected (Fig. 1c). Which of the two signatures reflects the actual valence and conduction band dynamics and how can we understand this apparent inconsistency taking into account well known physics such as band filling, bandgap renormalization and heating?

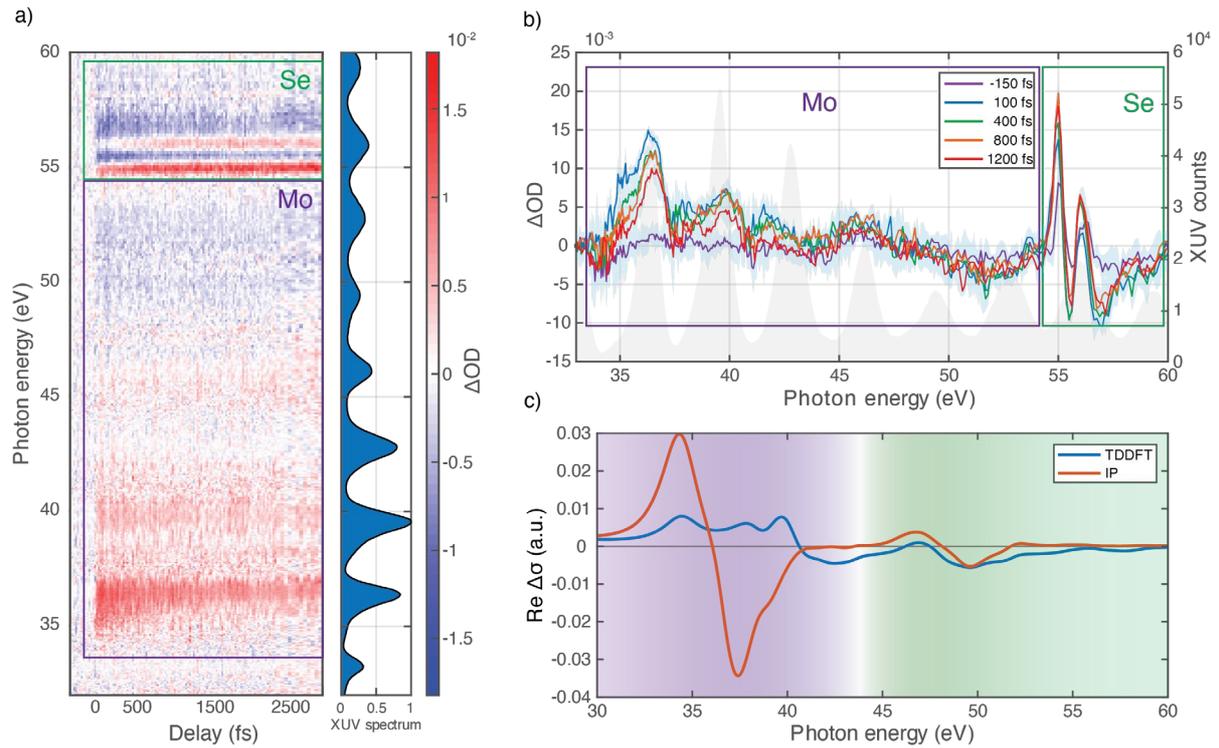

**Fig.2 a)** Attosecond transient absorption spectroscopy (ATAS) measurement of a few-layer $MoSe_2$ sample. The XUV probe response probed via the transitions from the Mo 4p states is marked in purple whereas Se 3d probed states are marked in green. **b)** Profile of the induced change in optical density at different probe times (shaded areas represent the standard deviation within the ±20 fs integration window). The transition metal Mo signal does not show any negative ΔOD around the bandgap, while the chalcogenide Se signal exhibits a clear negative ΔOD repeated by the *3d* core level spin orbit splitting. **c)** Independent particle (IP) and TDDFT calculations show the change in the real part of the optical conductivity. The IP calculation features a negative response for the Mo transition. Only TDDFT calculations including many body effects correctly reproduce the positive change for the Mo transition, thus highlighting the importance of localized screening due to electron localization into *d*-orbitals around the Mo atom after excitation. Spin-orbit splitting of the Se *3d* core level is not considered in these calculations.

The strongly energy-modulated, derivative-shaped features that appear in the Se response have been observed in previous ATAS measurements in semiconductors (see, e.g., [13,16,19]). A NIR pump induced increase of carriers in the CB manifests itself as a negative change of optical density (ΔOD) due to the reduced availability of unoccupied final states for the probe transition. The induced holes in the VB lead to more available final states and an associated positive optical density change. The combination of these two effects leads to the derivative-shaped structure with sharp positive ΔOD in the VB (missing electrons) and negative ΔOD in the CB (electrons excited from the VB). This spectral response and associated dynamics can be well described by a simple model assuming a Gaussian distribution of carriers in the conduction/valence band and the effects of band filling, bandgap renormalization and (thermal) red shift (see SI). A simple estimate based on carrier density of the induced bandgap renormalization results in ~160 meV renormalization [29,30], even though higher values up to 500 meV have been reported for TMDCs [31-33]. The described spectral response for the Se transition can be further confirmed with an independent particle model (IP) using time-dependent density functional theory (TDDFT) in Fig. 2c (see SI) [19].

The broad positive response in the Mo-specific absorption signal on the other hand does not indicate the increase in CB population which would be linked to a negative ΔOD. Only higher lying bands >47 eV above the probed Mo core level show a broad negative response (Fig. 3c). The states in this energy region are a mix of higher-lying bound and delocalized states, with the Mo *5s* contributing the most (see SI).

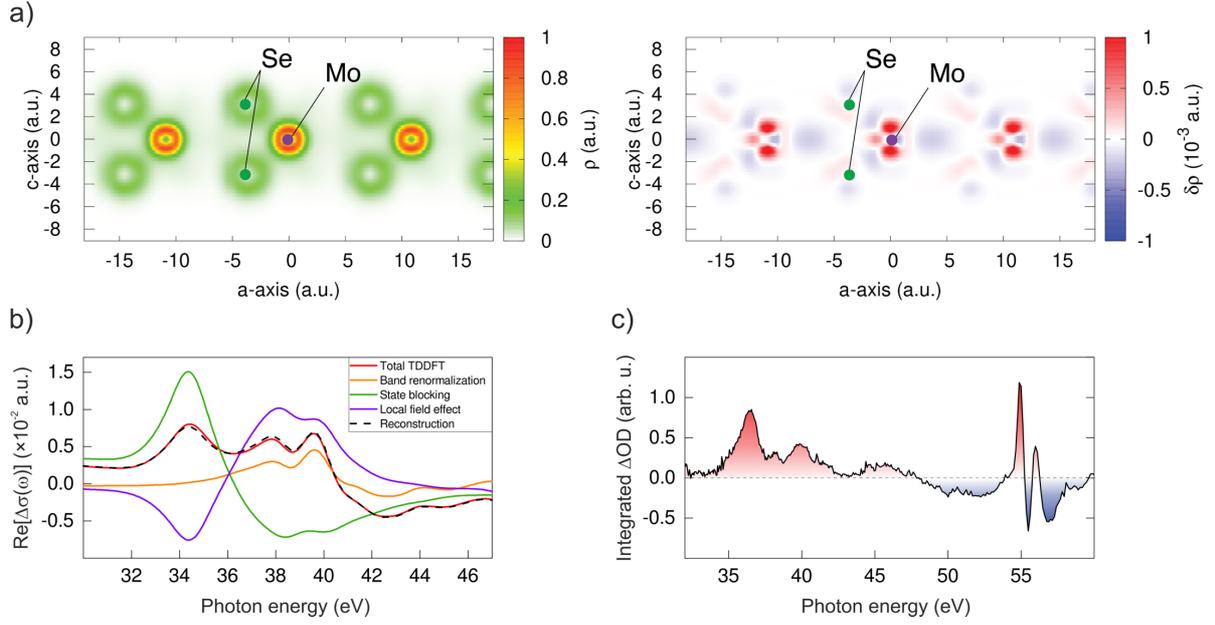

**Fig. 3 a)** Real-space electron distribution and change induced by excitation with the NIR pump pulse. The electrons strongly localize around the Mo site within the *d*-orbital (positive changes in this figure indicate an increased electron density). The densities are evaluated for a thermalized electronic state, corresponding to delays of several tens of femtoseconds after excitation. **b)** Decomposition of the TDDFT results in band filling, band renormalization and local screening modification contributions. The *d*-orbital localization results in a screening response that compensates the band filling effect and creates a broad overall positive change in absorption. **c)** ATAS signal integrated over delay from 50 fs to 4.3 ps. A negative response is also seen for higher lying states in the Mo signal at photon energies >47 eV which are dominated by Mo *5s* orbitals.

The simple model based on band filling, bandgap renormalization, and heating induced red shift that describes the signal observed for the Se probe transition completely fails to reproduce the measured spectral response for the Mo transition (see SI for more details). We therefore turn the focus to the different orbital configuration of the Mo and Se VB/CB contributions. Ab-initio calculations, in particular TDDFT with the adiabatic local density approximation (ALDA) and independent particle (IP) calculations (see SI), are used to verify the origin of the difference in the spectral response (Fig. 2c). The TDDFT calculations can qualitatively reproduce the spectral shape for both initial probe states. It should be noted that the calculated

Se response is located at a different energy compared to the experiment. This is due to the different core level energy of 45 eV instead of 55 eV found by TDDFT due to the lack of the core-hole effect in the ALDA. Furthermore, our TDDFT calculations do not consider spin-orbit splitting of the Se initial probe state. In contrast to the TDDFT, independent particle calculations do not include many-body effects, such as localized screening due to carrier localization. Figure 2c shows the comparison between the TDDFT and IP calculations, where the Mo transition shows the most prominent difference, while the two models agree qualitatively for the Se case. The IP calculation for the Mo transition exhibits a prominent increase and decrease in the optical conductivity, corresponding to the VB/CB respectively, rather than the broad positive response seen in the experiment. This difference between TDDFT and IP calculations indicates that the broad positive feature in the Mo response must be due to many-body effects.

To obtain further insights into the nature of these many-body effects, real space TDDFT calculations are performed to investigate the pump-induced charge redistribution. The results demonstrate a strong electron localization into the *d*-orbital around the Mo atom after the NIR pump pulse (Fig. 3a). A much weaker change with *p*-orbital symmetry is found around the Se atom. The *d*-orbital localization results in a strong local field effect with stronger screening as observed before in Ti metal [1]. However, it comes as a surprise that this *d*-orbital localization is taking place in the TMDC semiconductors and leaves the chalcogen response unaffected even though the transition metal *d*-orbital and chalcogen *p*-orbital overlap defines the band structure in the valence and conduction band.

A decomposition of the TDDFT calculated change of absorption around the molybdenum transition in MoSe$_2$ highlights the quantitative contributions of band filling, renormalization, and local screening modification to the overall signal (Fig. 3b). A strong contribution of the local screening effect to the spectral response is found. Its spectral shape originates from the presence of the giant resonance in molybdenum that increases its energy separation from the probe initial state in response to the screening due to the localization into Mo 4*d* orbitals, resulting in a shape that counteracts the band filling signal.

Based on this theoretical analysis, we assign the origin of the measured broad spectral feature of the Mo transition to the valence electron localization occurring around the 4*d* orbital of molybdenum. The negative ΔOD response of the higher lying bands probed by the Mo transition at photon energies between 47 and 54 eV shows further evidence for the screening effect most pronounced for *d*-orbitals (Fig. 3c). Remarkably, the Se response remains entirely unaffected by the electron localization, even though the CB is formed by an overlap of Mo 4*d* and Se 4*p* orbitals and the excited carrier population can be treated as an electron-hole liquid formed through global excitation from our NIR pump pulse (see SI). The latter aspect renders the strong localization around the Mo sites a counter-intuitive observation.

**Comparison with Ti and dynamical response**

Similar strong localization into transition metal *d*-orbitals as for the Mo response in MoSe$_2$ and an associated broad positive change in absorption was first observed for Ti metal [1]. In that study, it was shown that this is a universal feature for elementary *3d* transition metals. Here, we find that the *d*-orbital localization still dictates the behavior of the transition metal even when embedded into TMDC compounds. Furthermore a recent control experiment in Al has shown that such localization does not occur in an ordinary (non-transition) metal [21]. While the Se signal shows the typical ultrafast dynamics known from semiconductor materials, involving band filling, band renormalization, carrier thermalization and thermal energy shifts (see SI), the *d*-orbital localization dominating the Mo signal is a comparably long lasting effect. The broad positive ΔOD feature in our ATAS data decays exponentially with a time constant on the order of 800 fs (Fig. 4a).

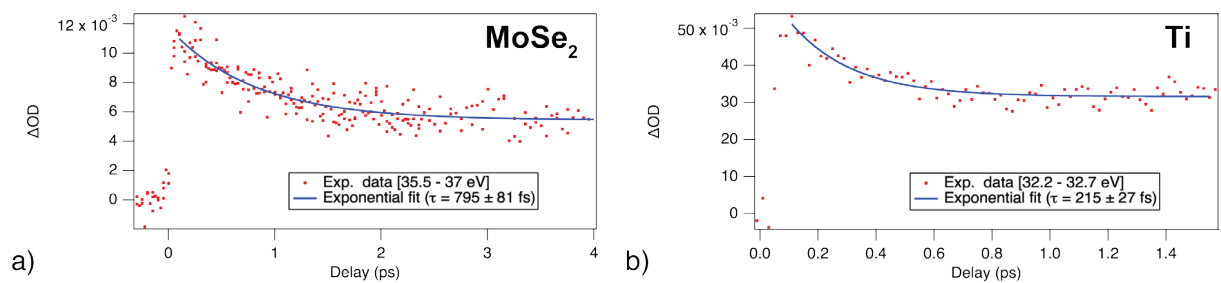

**Fig 4. a)** Decay of the main positive ΔOD signal in MoSe$_2$ in the photon energy band from 35.5 to 37 eV. An exponential fit reveals a decay time of ~800 fs while the overall signal remains positive for many picoseconds. **b)** Decay of ΔOD signal in vicinity of giant resonance of Ti metal (see [1] for details). The data has been integrated in a photon energy interval of 32.2 to 32.7 eV. An exponential fit yields a decay time of ~220 fs.

In Ti metal, on the other hand, the main positive feature around the Fermi energy decays within ~220 fs (Fig. 4b). In another data set covering higher photon energies, we find 6 eV above the Fermi energy a negative ΔOD signal that appears about 75 fs after the pump pulse (see SI). While the latter Ti data has relatively poor statistics, a comparison with theory leads us to speculate that the appearance of the negative signal can be attributed to a fast decay of the *d*-orbital localization, revealing the state-filling part of the signal from the Ti *3d* and *4s* orbitals (see SI and [1]). Furthermore, we can assume that the electron localization in Ti decays much faster than the localization around Mo in the semiconductor MoSe$_2$ due to the much higher electron-electron scattering rates in the metal compared to a material with a bandgap.

**Conclusion**

In conclusion, we present the ultrafast measurement of excited state carrier localization in semiconductors with transition metals such as MoSe$_2$. Through simultaneous element and carrier specific probing of the excited states, we identify the real space carrier localization right after excitation around molybdenum. While the spectral response of the transition metal is dominated by screening effects with ultrafast *d*-orbital localization, the selenium is not affected by this collective response and can be described by an independent particle model with band filling, thermalization and lattice heating. This can be considered surprising given that the NIR-pump excited carriers are well above the Mott-transition. This means the overall dynamics in such materials with transition metal elements cannot be described with an independent particle model. This may have implications for the applicability of ubiquitous effective mass approximation for describing such semiconductors. The localization effects appear rapidly with the pump excitation on a few-femtosecond scale but appear to have a relatively long lifetime in semiconductors compared to the transition metal Ti. Furthermore, the real space carrier localization adds an additional degree of freedom to the toolbox of engineering of opto-electronic materials, in addition to previously studied valley-selectivity and spin-momentum locking. Understanding the effects of the observed real space carrier localization is the first step towards control of excited states on an element specific level in novel 2D energy materials. The qualitatively different ultrafast carrier dynamics observed for the transition metal Mo with the bonding *d*-orbitals and the chalcogen Se with the bonding *p*-orbitals forming both the conduction and valence bands demonstrates the need for element and carrier specific measurements also for heterostructure 2D systems where interlayer charge transfer occurs, a crucial process to advance TMDCs for future device applications. The present results are expected to be general and would be applicable to the whole family of homo and hetero TMD layered structures, including Moiré heterostructures.

## Methods

**Laser parameters** The pulse intensity is calculated based on the pulse width of 15 fs measured with spectral phase interferometry for direct electric-field reconstruction (SPIDER[34]) and a pulse energy of 1 μJ (1 μW average power at 1 kHz pulse repetition rate). The focal spot has dimensions of 80 μm (horizontal) by 60 μm (vertical). According to our SPIDER measurements, 90% of the pulse energy are contained in the main pulse. Measurements of the spatial beam profile show that also 90% of the energy is focused into the central spot in the focal plane. We therefore estimate an intensity of $5.8*10^{11}$ W/cm$^2$ ± $1*10^{11}$ W/cm$^2$.

**Delay zero calibration and temporal resolution** Delay zero calibration for the overlap between the NIR and XUV pulse train was obtained by acquiring a full RABBITT (reconstruction of attosecond beating by interference of two-photon-transitions [35]) data set in argon. The envelope of the side bands that appear in the region of pulse overlap are fitted to obtain delay zero and gain information about the cross-correlation width. The latter indicates a temporal resolution of our experiments of 10.3 ± 0.3 fs. From the oscillation period, we extract a center photon energy of our NIR pump pulse of 1.54 eV.

**Materials** The MoSe$_2$ samples are multilayer CVD grown MoSe$_2$ films (6Carbon Technology), specified with a typical photo-luminescence signal centered at 800 nm. The samples are provided on 30 nm thick silicon nitride membranes from Agar Scientific (AGS171-1).

**Acknowledgments** The authors kindly acknowledge the experimental contribution of Mikhail Volkov for the data analyzed and displayed in Fig. 4b. This research was supported by the NCCR MUST, funded by the Swiss National Science Foundation, by the Swiss National Science Foundation (SNSF) project 200020_200416 and by JSPS KAKENHI Grant Number JP20K14382. Z. Schumacher gratefully acknowledges the support by the ETH Zurich Postdoctoral Fellowship Program. This project has received funding from European Union's Horizon 2020 under MCSA Grant No 801459, FP-RESOMUS. This project was supported by the European Research Council (ERC-2015-AdG694097), the Cluster of Excellence 'CUI: Advanced Imaging of Matter' of the Deutsche Forschungsgemeinschaft (DFG) - EXC 2056 - project ID 390715994, Grupos Consolidados UPV/EHU (IT1249- 19), partially by the Federal Ministry of Education and Research Grant RouTe-13N14839, the SFB925 "Light induced dynamics and control of correlated quantum systems", by The Flatiron Institute, a division of the Simons Foundation.

# Supplementary Information

# Ultrafast electron localization and screening in a transition metal dichalcogenide

Z. Schumacher[1], S. A. Sato[2,3], S. Neb[1], A. Niedermayr[1], L. Gallmann[1,*], A. Rubio[3,4], U. Keller[1]

[1]Department of Physics, ETH Zürich, 8093 Zürich, Switzerland

[2]Center for Computational Sciences, University of Tsukuba, Tsukuba, Ibaraki 305-8577, Japan

[3]Max Planck Institute for the Structure and Dynamics of Matter and Center for Free Electron Laser Science, 22761 Hamburg, Germany

[4]Center for Computational Quantum Physics (CCQ), Flatiron Institute, New York 10010, USA

*Corresponding author. gallmann@phys.ethz.ch

## S1 Static absorption measurement

We performed static absorption measurements of our few-layer MoSe$_2$ films (6Carbon Technology) on 30 nm thick silicon nitride membranes from Agar Scientific (AGS171-1). For comparison, we also recorded static absorption data of mono layer samples from the same supplier. All data was taken in our attosecond transient absorption spectroscopy (ATAS) setup. The comparison is shown in Figure S1 below. The static absorption was measured in the range of 29 to 42 eV, where our setup delivers the highest flux. The data is limited to discrete energy intervals due to the harmonic structure of our XUV spectra. A static absorption measurement with similar fidelity around the Se edge was not possible due to the lower XUV count rate and the generally weaker absorption of Se at the relevant energy. The transient signal used for the ATAS scans overcomes these issues by fast averaging methods and the fact that only the difference in absorption between NIR excitation and ground state is measured. The strong signal from the excited sample further benefits the signal quality for the transient signal, while we cannot take advantage of this effect for the static measurements.

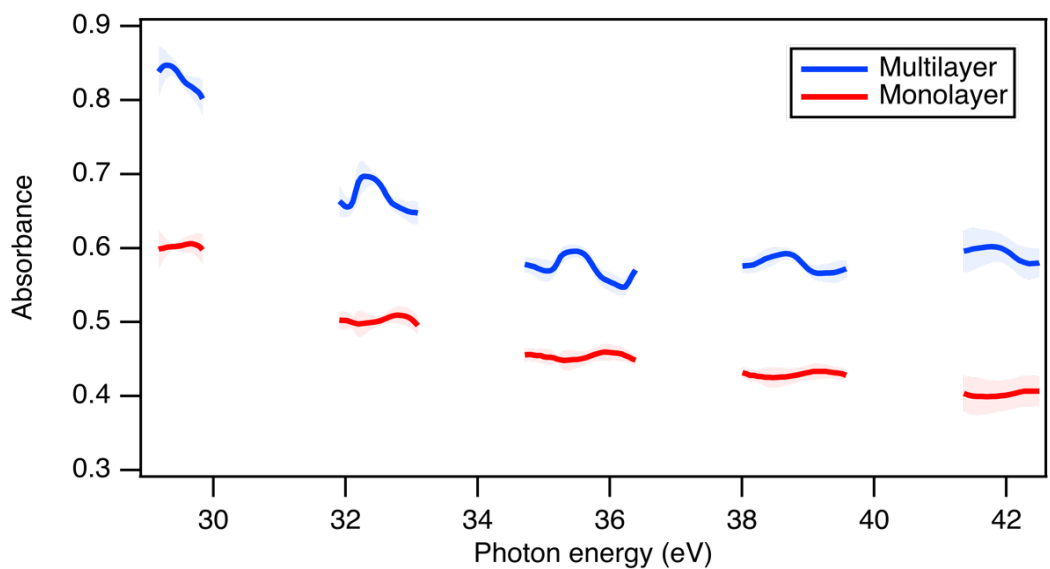

**Fig. S1.** Static absorption measurement of mono- and multilayer MoSe$_2$. The discrete photon energy intervals are due to the harmonic spectrum of the probe XUV spectrum. The shaded areas represent the error bars (standard deviation over 3600 acquired sample and reference transmission spectra).

## S2 Estimates of sample excitation

### S2.1 Number of excited carriers

The number of excited carriers per pulse is obtained by using a model that was previously applied to other 2D materials [1]. The carrier area density per pulse (cm$^{-2}$) is given by:

$$n_0 = \frac{P * \sigma}{A * F_{rep} * E}$$

with the average power P, the focal spot size A, the pulse repetition rate $F_{rep}$, the photon energy E, and the absorptance $\sigma$. A value of $\sigma = 0.03$ is used based on the dielectric constant (4) and sample geometry of MoSe$_2$. We obtain a carrier density of $1.95*10^{14}$ cm$^{-2}$ ± $0.75*10^{14}$ cm$^{-2}$. The error bar is mainly due to uncertainties in the sample position within the focal spot. This carrier density is 1-2 orders of magnitude above the expected Mott density [2].

A similar estimate of carrier volume density can be obtained by using the fluence and absorption coefficient of MoSe$_2$ as presented by Zürch, et al. in 2017 [3]. Here, the fluence F, the central frequency ν of the laser pulse, the absorption coefficient α, and the sample thickness is used. We find the carrier density therefore as follows:

$$N_e \cong \frac{F}{h\nu d}[1 - e^{-\alpha d}]$$

We estimate a carrier density in the range of $2.1*10^{21}$ cm$^{-3}$ ± $0.8*10^{21}$ cm$^{-3}$ for an estimated sample thickness of 2 nm (3-4 layers). A sample thickness variation has a negligible effect compared to the uncertainty in fluence due to the sample position. A thickness of 3 nm would change the maximum carrier density by $0.023*10^{21}$ cm$^{-3}$.

## S2.2 Estimation of bandgap renormalization

For the estimation of the bandgap renormalization, the method presented by Bennett et al. [4] is used:

$$\Delta E_g = \left(\frac{e}{2\pi\epsilon_0\epsilon_s}\right)\left(\frac{3}{\pi}\right)^{\frac{1}{3}} N_e$$

For MoSe$_2$ we use $\epsilon_s \approx 21.8$ (4) for the relative static dielectric constant, and the concentration of free electrons $N_e$ calculated above. A bandgap renormalization in the range of -187 meV to -143 meV, taking the uncertainty in carrier density into account, is found.

# S3 Dynamics in the Se response

In the following, we discuss the dynamics observed in the signal obtained from the probe transition via the Se core level. Our data confirms observations that were previously reported for the closely related material $MoTe_2$ [5].

In figure 2b) of the main manuscript, lineouts of the spectral response for different delay times are shown. The prominent Mo peak above 35 eV decays with increasing time delay, while the response from the selenium transition is more complex. We concentrate here on the lower of the two derivative-shaped replica in the Se signal that arise from the two spin orbit split Se *3d* core states. At first it might appear as if the VB signal strength increases (first positive peak below 55 eV), suggesting an increased hole population in the valence band. However, we need to consider the simultaneously occurring effects of carrier thermalization and lattice heating to gain full insight. We therefore use a decomposition method to identify the contribution of state-blocking, broadening and redshift of the absorption edge [3] to the measured transient spectrogram (see Fig. S2). Details on the decomposition approach are given below.

The band filling signal in Figure S2b shows the VB peak to narrow and move to higher energies, indicating a thermalization of the carriers to the top/bottom of the VB/CB, as expected for the initial non-thermal distribution. Simultaneously, we see a decay of the broad population of electrons in the conduction band within the first picosecond by the decrease of the amplitude of the higher CB (arrow above 56.5 eV). The broadening and red shift (Fig. S2c) leads to more absorption below the absorption edge and results in the overall increase of absorption in the VB as seen in the measurement. From these observations we can draw two important conclusions. First, carriers in the VB/CB thermalize within 800 fs (shift in energy of the VB peak) and recombine, as seen by the band filling signal. Second, the apparent initial increase in VB signal strength is due to a superposition of multiple effects and, thus, to gain the full insight into the carrier dynamics from the ATAS trace, a decomposition needs to be performed to disentangle the various competing processes similar to previous results [3,5].

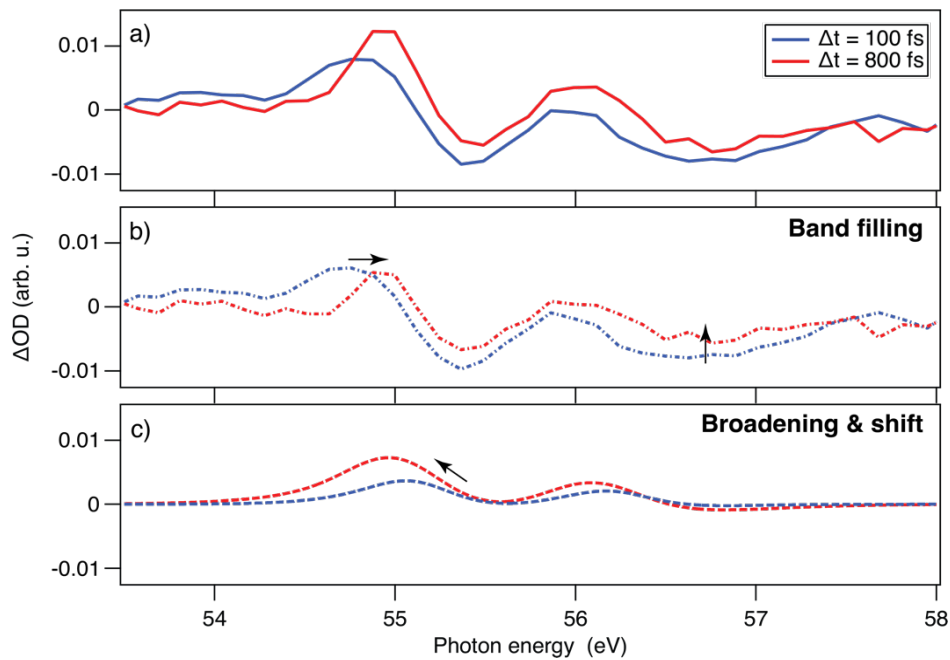

**Fig. S2** Evolution of the optical density change induced after photo excitation for Se core level to VB/CB. **a)** Measured change at two different pump-probe delays. **b)** The band filling and **c)** broadening & red shift contribution to the overall signal. Due to lattice heating, the latter contribution leads to an increase of the signal at 800 fs, resulting in an overall increase of the observed DOD at 55eV.

## S3.1 Decomposition of ATAS signal

The ATAS trace can be modelled using a static absorption spectrum, which is red shifted and broadened as well as combined with a state-blocking signal. We follow an adapted version of this reconstruction process, initially presented by Zürch et al. [3], as outlined below.

While it is preferred to use measured static absorption spectra for the input of the reconstruction, we must rely on calculated static spectra since the signal to noise of our static absorption data in the photon energy region of the Se transition does not allow for a robust decomposition of the signal from our thin $MoSe_2$ samples. We, therefore, calculate the static absorption spectrum for the Se transition based on the spin-orbit split core hole positions of 55.2eV and 56.3 eV, respectively. The positions of the core hole energies are obtained by using the zero ΔOD crossing in the ATAS scan as the center between the VB and CB position for both pairs of positive and negative absorption bands in the Se signal. The values obtained with this method deviate slightly from literature values for Se core hole energies. The deviation could be due to the substrate affecting the MoSe$_2$ sample, The absorption spectrum is then obtained with the following model functional form and $\Delta E = 0.2\ eV$:

$$A_{\text{static}}(E) = \frac{1}{1+\exp\left(-(E-E_{\text{corehole}_1})/\Delta E\right)} + \frac{1}{1+\exp\left(-(E-E_{\text{corehole}_2})/\Delta E\right)}$$

This static spectrum is then shifted and broadened by a convolution (denoted as $*$ below) as follows:

$$\Delta A_{\text{shift}}(E) = A_{\text{static}}(E) - A_{\text{static}}(E + E_{\text{shift}})$$

$$\Delta A_{\text{broad}}(E) = (A_{\text{static}}(E) * \exp\left(-E^2/(2\sigma^2)\right)) - A_{\text{static}}(E)$$

The initial band filling signal is calculated based on a gaussian distribution of holes/electrons in the VB/CB resulting in an increased/decreased absorption. Each distribution is calculated based on the following functional form (see Zürch et al. [3], SI. Eq. 7) :

$$\Delta A_{\text{band filling}}(E) = \exp\left(-\frac{4\log(2)(E-E_{VB})^2}{BW_{\text{NIR}}^2}\right)$$

with $E_{VB}$ the position of the VB/CB in the absorption spectrum and $BW_{\text{NIR}}^2$ the bandwidth of the NIR excitation. For the latter, an initial starting value of 0.4 eV was used.

Two exemplary reconstructions at fixed delay times are shown in Fig. S3 and S4. The transient response is averaged for +/- 30 fs around 100 and 800 fs, respectively. For these two time delays, we obtain reasonable agreement with the experimental data for a broadening of 0.3 and 0.6 eV for 100 and 800 fs, respectively. The observed shift increases from 0.2 eV at 100 fs to 0.4 eV at 800 fs.

We employed the same decomposition approach for the molybdenum transition at 34 to 44 eV. However, the reconstruction did not yield a good agreement with the experimental data (see fig. S5). This is further evidence that the signal from the Mo transition is dominated by the screening dynamics, which is not taken into account in the simple model underlying this decomposition.

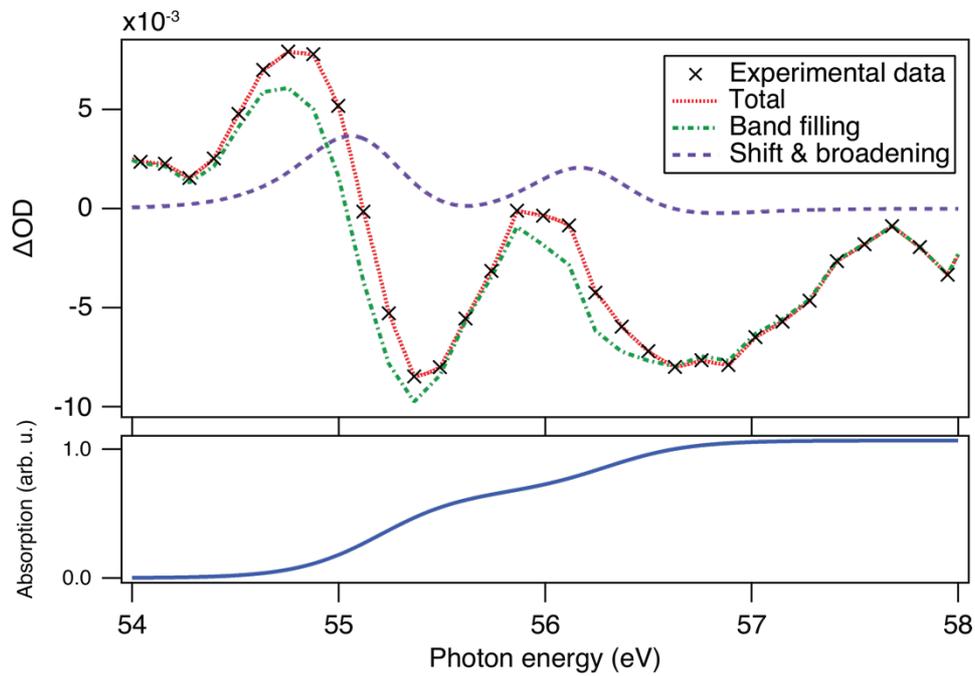

**Fig. S3.** Decomposition of ATAS trace for the observed selenium transition at 100 +/- 30 fs. The shift and broadening of the static absorption (bottom panel) is shown as a dashed purple line, with the band filling as a dash-dotted green line and the measured signal as black crosses. The red dotted trace shows the sum of the decomposed contributions.

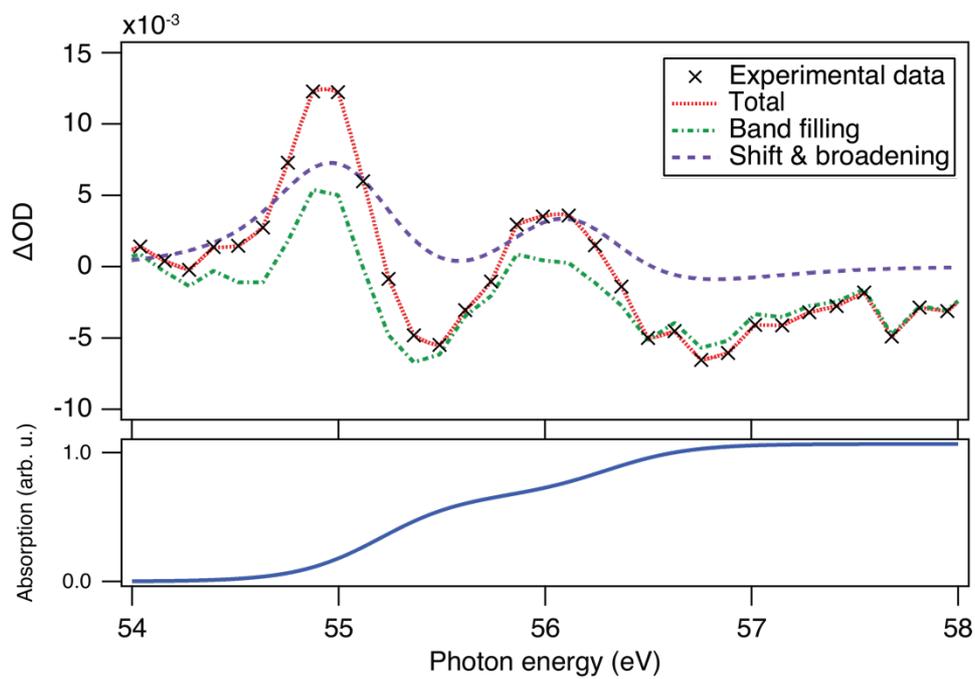

**Fig. S4.** Same as Figure S3, but for a pump-probe delay of 800 +/- 30 fs. The color code is identical to Figure S3.

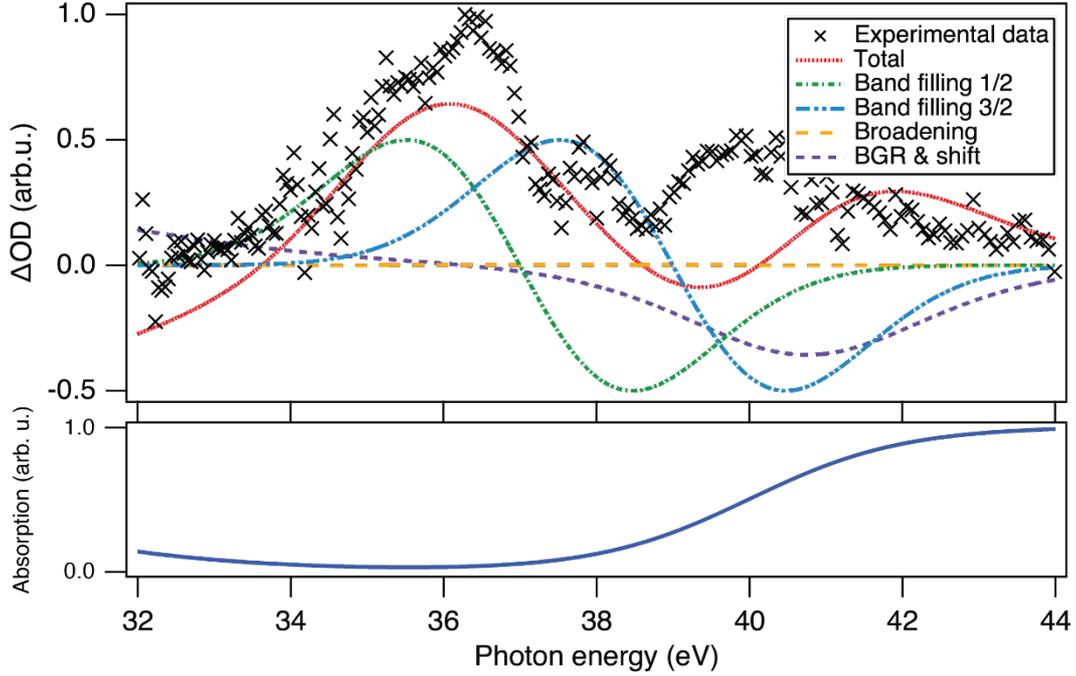

**Fig. S5.** Attempted decomposition of ATAS trace for the observed molybdenum transition at a delay of 100 ± 30 fs. The total calculated signal (red dotted) does not agree well with the measured data (black crosses). This is further evidence that screening dynamics dominate the Mo response, which are not included in this model. The static absorption is shown in the bottom panel.

## S4 Theory

### S4.1 First-principles electron dynamics calculation with time-dependent density functional theory

We analyze the microscopic electron dynamics in the material with first-principles electron dynamics calculations based on time-dependent density functional theory [6]. Because the calculation method has been described in detail elsewhere [7], here, we only briefly describe the method for electron dynamics in MoSe$_2$. For practical calculations of electron dynamics, we solve the following time-dependent Kohn-Sham equation for electron orbitals:

$$i\hbar\frac{\partial}{\partial t}\psi_{bk}(\mathbf{r},t) = \left[\frac{(\mathbf{p}+\mathbf{A}(t))^2}{2m_e} + \hat{v}_{ion} + v_H(\mathbf{r},t) + v_{XC}(\mathbf{r},t)\right]\psi_{bk}(\mathbf{r},t), \quad (1)$$

where $b$ is a band index, $\mathbf{k}$ is the Bloch wavevector, and $\mathbf{A}(t)$ is a time-dependent vector potential, which is related to the external laser electric field as $\mathbf{E}(t) = -d\mathbf{A}(t)/dt$. The equation of motion, Eq. (1), contains the electron-ion interaction $\hat{v}_{ion}$, the Hartree-potential $v_H(\mathbf{r},t)$, and the exchange-correlation potential $v_{XC}(\mathbf{r},t)$. In this work, we employ the adiabatic local density approximation (ALDA) for $v_{XC}(\mathbf{r},t)$ [8]. To describe the electron-ion interaction, we employ the norm-conserving pseudopotential for both Mo and Se. For Mo, *4s*, *4p*, *4d*, and *5s* electrons are treated as valence. For Se, we employ two sets of pseudopotentials. The first one is used to describe the excitations from semi-core states of Se, treating *3d*, *4s*, and *4p* electrons as valence. The second one is used to describe the valence electron dynamics, treating *4s* and *4p* electrons as valence.

To analyze the optical property of the material, we compute the optical conductivity of the monolayer MoSe$_2$ with the first-principles electron dynamics calculation. For this purpose, we employ the real-time linear response calculation, by evaluating the electron dynamics with Eq. (1) under an impulsive distortion, $\mathbf{E}(t) = \mathbf{e}_x E_0 \delta(t)$. By using the induced current density $\mathbf{J}(t)$, the optical conductivity can be evaluated as

$$\sigma(\omega) = \frac{1}{E_0}\int_0^\infty dt\, \mathbf{e}_x \cdot \mathbf{J}(t) e^{i\omega t - \gamma t},$$

where $\gamma$ is the damping parameter, which is set to 1.0 eV in this analysis.

To elucidate the modulation of the optical property of MoSe$_2$ by laser excitation, we compute the optical conductivity of the monolayer MoSe$_2$ with a hot-electron state, where the electron temperature is set to 0.5 eV, while the lattice remains cold [9]. The result of the optical conductivity modulation with the electron temperature increase is shown in Fig. 2c) of the main text. Furthermore, the result with the independent-particle (IP) approximation, where the time-dependence of the Hartree and exchange-correlation potentials is ignored, is also shown in Fig. 2c) to demonstrate the important role of many-body effects.

We further analyze the microscopic mechanism of the modification of the optical absorption of MoSe$_2$ by decomposing the transient conductivity into the three components [7]:

$$\Delta\sigma(\omega) = \Delta\sigma^{BR}(\omega) + \Delta\sigma^{SB}(\omega) + \Delta\sigma^{LF}(\omega),$$

where $\Delta\sigma^{BR}(\omega)$ is the contribution from the band renormalization, $\Delta\sigma^{SB}(\omega)$ is the contribution from the state-filling, and $\Delta\sigma^{LF}(\omega)$ is the local-field effect. The decomposed results and reconstructed data are shown in Fig. 3b) in the main text. A detailed description of the decomposition approach is provided in previous work [7].

To obtain further insight into the microscopic electron dynamics in MoSe$_2$, we compute the electron density dynamics under the irradiation of laser fields. We solve the time-dependent Kohn-Sham equation, Eq. (1), by employing a laser pulse. Here, we set the full duration of the laser pulse to 20 fs, the mean photon energy to 1.55 eV, and the peak field strength to 20 MV/cm. The initial electron density $\rho(\boldsymbol{r}, t = 0)$ is shown in Fig. 3a) in the main text along with the electron density difference $\delta\rho(\boldsymbol{r})$ induced by the laser excitation. Here, the electron density difference $\delta\rho(\boldsymbol{r})$ is evaluated as the average density difference, $\delta\rho(\boldsymbol{r}, t) = \rho(\boldsymbol{r}, t) - \rho(\boldsymbol{r}, t = 0)$ over 10 fs after the laser irradiation.

### S4.2 Orbital character of higher lying bands in Mo signal

To analyze the electronic structure of matter, we employ the density functional theory using the Quantum ESPRESSO package [10]. Figure S6 displays the data shown in Fig. 1c) of the main manuscript over a wider energy range. The higher lying states that give rise to the broad derivative shaped feature at photon energies >43 eV in the Mo signal consist of a wide range of bound and delocalized states. The most dominant contribution comes from the Mo *5s* orbital.

This is also consistent with previous observations in MoS$_2$ using resonant photoemission [11]. Clearly, *d*-orbital contributions play a negligible role in this energy window.

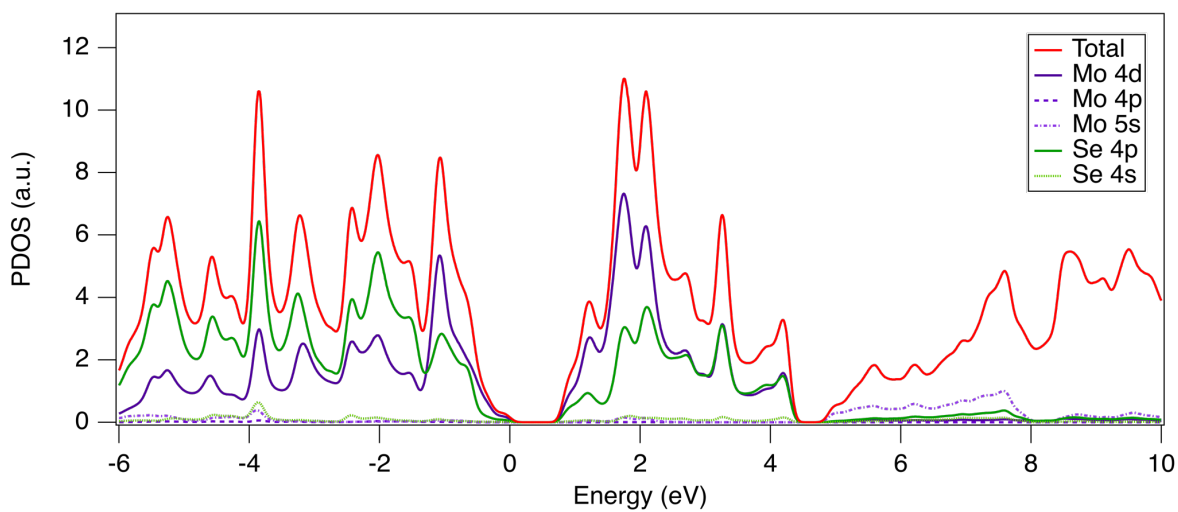

**Fig. S6** Projected density of states of bulk MoSe$_2$. This is an extended version of Fig. 1c, plotted over a wider energy range. In the band at energies of ~5 to >10 eV, contributions from *d*-orbitals play a minor role. Instead the Mo *5s* orbital is the most dominant constituent among a multitude of other bound and delocalized states.

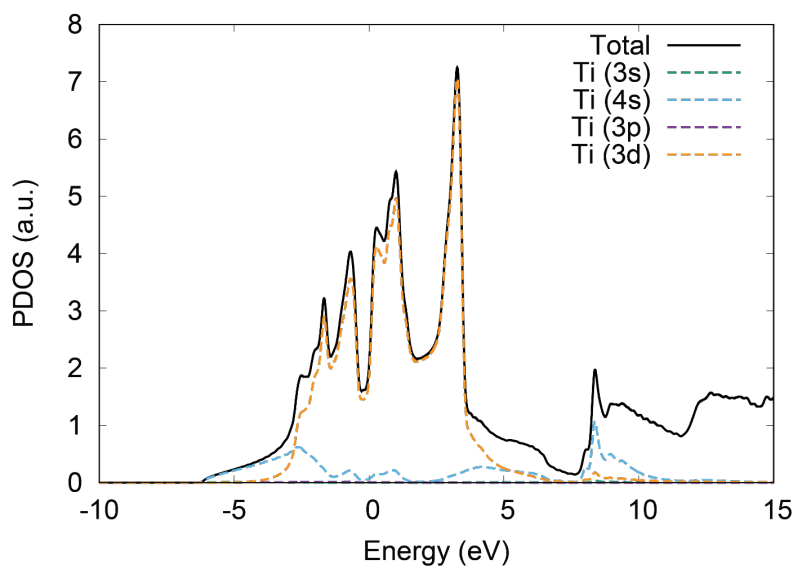

**Fig. S7** Projected density of states of titanium. In the region of the giant resonance around the Fermi energy, the density of states is dominated by *3d* orbitals. About 6 eV above the Fermi energy, *3d* and *4s* states contribute with comparable strength.

## S4.3 Projected density of states for Titanium

Figure S7 displays the projected density of states for titanium. The energy scale is referenced to the Fermi energy, which is located in the so-called giant resonance of titanium. At those energies, the dominant contribution to the density of states comes from *3d* orbitals. Only at about 5-6 eV above the Fermi energy, *3d* and *4s* orbitals contribute with similar strength.

## References SI